\documentclass[preprintnumbers,amsmath,amssymb,aps,preprint]{revtex4}

\usepackage{graphicx}
\usepackage{multirow}
\usepackage{tabularx}

\newcommand{\bee}{\begin{equation}}
\newcommand{\ee}{\end{equation}}
\newcommand{\beea}{\begin{eqnarray}}
\newcommand{\eea}{\end{eqnarray}}
\newcommand{\gfive}{\gamma_5}

\def\Tr{{\rm Tr}}

\begin{document}

\title{Quark condensate in one-flavor QCD}

\author{Thomas DeGrand, Roland Hoffmann, Zhaofeng Liu}
\affiliation{
Department of Physics, University of Colorado,
Boulder, CO 80309 USA
}

\author{Stefan Schaefer}
\affiliation{
NIC, DESY Zeuthen,
Platanenallee 6,
D-15738 Zeuthen,
Germany
}

\begin{abstract}
We compute the condensate in QCD with a single quark flavor using
numerical simulations with the overlap formulation of lattice fermions. The condensate
is extracted by fitting the distribution of low lying eigenvalues of the Dirac operator
in sectors of fixed topological charge to the predictions of Random Matrix Theory.
Our results are in excellent agreement with estimates from the orientifold large-$N_c$ expansion.
\end{abstract}

\maketitle

\section{Introduction}
Very few analytic techniques are available to study nonperturbative properties of QCD.
Of these, the most prominent are large-$N_c$ expansions.
Recently, Armoni, Shifman, and 
Veneziano~\cite{Armoni:2003fb,Armoni:2003gp,Armoni:2003yv,Armoni:2004ub}
 suggested a new large-$N_c$ expansion
with some remarkable features.
In contrast to the 't Hooft large-$N_c$ limit~\cite{'tHooft:1974hx}
 ($N_c\rightarrow \infty$, $g^2 N_c$ and $N_f$ fixed,
with quarks in the fundamental representation of $SU(N_c)$), quarks are placed in the
two-index antisymmetric representation of $SU(N_c)$. Now in the 
$N_c\rightarrow \infty$, $g^2 N_c$ and $N_f$ fixed limit of QCD, quark effects are not
decoupled, because there are as many quark degrees of freedom as gluonic ones, $O(N_c^2)$
in either case. In Ref.~\cite{Armoni:2004ub} the
authors have argued that a bosonic sector of ${\cal N}=1$ super-Yang-Mills theory
is equivalent to this theory in the large-$N_c$ limit.
(The proof of this connection has recently been extended to lattice regularized theories
by Patella~\cite{Patella:2005vx}.)
 The large-$N_c$ QCD-like theory
is called ``orientifold QCD.''

For $N_c=3$, orientifold QCD is equivalent to QCD with a single quark flavor
in the fundamental representation of $SU(3)$. This equivalence even extends to the
first and second  terms in the $\beta$ function and in the lowest order
anomalous dimension for the running quark mass (or quark condensate).
This means that nonperturbative quantities computed in super-Yang-Mills theory
can be related to corresponding ones in one-flavor QCD, up to $1/N_c$ effects.

In a recent
 paper~\cite{Armoni:2003yv}, Armoni, Shifman and Veneziano estimate the quark condensate
in one-flavor QCD from the value of the gluino condensate in SUSY Yang-Mills. 
They find (with our sign conventions)
\bee
\Sigma = \{0.014, \ 0.021, \ 0.028\} \ {\rm GeV}^3
\label{eq:asv}
\ee
in the $\overline{MS}$ scheme at $\mu=2$ GeV. The spread of values gives their estimate
of $1/N_c$ corrections. This corresponds to $\Sigma^{1/3}$ of
240 to 300 MeV. They estimate values of $\Sigma$ from the Gell-Mann, Oakes, Renner relation
and from an interpolation of lattice data, which are in good agreement with Eq.~(\ref{eq:asv}).
However, a direct lattice calculation of the condensate in $N_f=1$ QCD
would give a more reliable comparison. Such a calculation would be important
to researchers studying the orientifold theory, for it would indicate
the size of  $1/N_c$ corrections to calculated quantities. This lattice calculation
 we now provide.

In the literature, there are different but related quantities  called the quark condensate. 
Often, the expression refers to $\langle \bar q q \rangle$, which in a lattice simulation, 
as well as in the continuum,
is a function of the quark mass and the volume in which the quark fields are defined.

In this paper, however, we attempt to extract the low-energy constant $\Sigma$, i.e. 
a parameter of the low-energy effective theory. 
The case of one flavor is a bit special: the chiral symmetry is anomalous, 
there is a massive pseudoscalar, the $\eta'$, and no Goldstone bosons.
The $\Sigma$ which we are about to extract is therefore not an order parameter
of spontaneous chiral symmetry breaking. However, there still exists a well defined 
low-energy description of $N_f=1$ QCD.
It has been worked out by Leutwyler and Smilga~\cite{Leutwyler:1992yt} to which we refer 
the reader for details. 
They show that up to terms of order $m^2V$  the partition function is 
\bee
Z=\exp\big\{\Sigma V  {\rm Re}(me^{-i\theta}) \big\}
\ee
with $\theta$ the vacuum angle. 
$\Sigma$ is the infinite volume zero quark mass limit of $-\langle \bar q q \rangle$
at $\theta=0$.

The lattice calculation has four parts:

First, we must describe how to perform simulations with one dynamical quark flavor.
These require the use of overlap fermions, for which algorithms have only recently
been developed~\cite{Fodor:2003bh}.
 Regularities in the spectrum of overlap fermions allow us to  use
 a version of the Hybrid Monte Carlo algorithm with chiral pseudofermions
to give us an exact algorithm for any number of flavors~\cite{Bode:1999dd,ouralg}.
The use of a lattice action with exact chiral symmetry means that observables (like the
anomaly) are not contaminated by explicit chiral symmetry breaking effects from the discretization.

Since direct measurements of $\langle \bar q q \rangle$ are influenced by finite volume
and non-zero quark mass, we  use another quantity whose dependence
on the condensate is known.
For this we choose the low eigenvalues of the Dirac operator, measured in sectors of
 fixed topology
$\nu$ in a simulation volume $V$.
The probability distribution of individual eigenvalues $\lambda_n$ is given by 
Random Matrix Theory (RMT)~\cite{Shuryak:1992pi,Verbaarschot:1993pm,Verbaarschot:1994qf}
as a function of the dimensionless quantity $\lambda_n \Sigma V$,
which depends parametrically
on the combination $m_q \Sigma V$ and, of course, the number of flavors $N_f$.
(Formulas for the special case of $N_f=1$ can be found in  Ref. \cite{Leutwyler:1992yt}.)
We use the specific method and predictions from
 Refs.~\cite{Damgaard:2000qt,Damgaard:2000ah}, where all the relevant formulas are
displayed.

Third, we need a lattice spacing to convert the dimensionless lattice-regulated condensate
to a dimensionful number. 
The calculation of Ref.~\cite{Armoni:2003yv} uses a $\Lambda$ parameter to set the
scale. Lattice calculations have not done that for many years, since typically
it is hard to extract a $\Lambda$ parameter from simulation data and because
most observed quantities in fact are not very sensitive to it.
Instead, it is customary to set the scale with
some spectral quantity, which can be computed
directly in a simulation. Possibilities include the masses
of various mesons or the string tension.
We follow common lattice practice and
 obtain the lattice spacing through the Sommer parameter $r_0$~\cite{Sommer:1993ce},
which is defined through the force,
\bee
-r^2 \frac{\partial V(r)}{\partial r}\Big|_{r=r_0} = 1.65\ .
\label{above_equation}
\ee
We could alternatively use the  string tension, but $r_0$ is less noisy.

The prediction of  Eq.~(\ref{eq:asv}) is a number in GeV. Since $N_f=1$ QCD
has no physical realization,  its overall scale is unknown. 
We will convert our dimensionless number into a dimensionful one using the
real world value of $r_0=0.5~{\rm fm}$, even though its ratio to other observables is almost
certainly $N_f-$dependent.

Finally, we need a matching factor, to convert the lattice-regulated condensate to its
$\overline{MS}$ value. We do this using
 the Regularization Independent scheme~\cite{Martinelli:1994ty}.

In Sec. \ref{sec:method} we describe all the ingredients of the lattice
calculation. Our results are summarized in Sec. \ref{sec:results}
and we conclude in Sec. \ref{sec:conclude}.

The only previous calculation of the condensate for $N_f=1$ we are aware of
is that of Ref.~\cite{Damgaard:2000qt}: the authors only quote $\Sigma a^3$.
The calculation was done deep in the strong coupling limit with a single
flavor of staggered fermions. Flavor symmetry was so badly broken that
the four tastes act as a single physical flavor.

\section{Lattice Methodology \label{sec:method}}

\subsection{Lattice action and simulation parameters}

Our simulations are performed with overlap fermions~\cite{Neuberger:1997fp,Neuberger:1998my}.
This discretization of the Dirac operator preserves exact chiral symmetry
at nonzero lattice spacing via the Ginsparg-Wilson relation~\cite{Ginsparg:1981bj}.
The massless operator is
\bee
D =D_{ov}(m=0)= R_0 \left[ 1+\gfive \epsilon(h(-R_0))\right]
\label{eq:Dov}
\ee
with $\epsilon(h)=h/\sqrt{h^2}$ the sign function of the Hermitian kernel
operator $h=\gfive d$ which is taken at  negative mass $R_0$.
It has a spectrum consisting of chiral modes with real eigenvalues
(at $\lambda=0$ and $2R_0$) and nonchiral modes
which are paired complex conjugates, $\lambda$ and $\lambda^*$.
The squared Hermitian overlap operator
$H^2=(\gamma_5 D)^2=D^\dagger D$
commutes with $\gamma_5$ and therefore its eigenvectors can be chosen with
definite chirality.  Because of the Ginsparg-Wilson relation, to each eigenvalue
$|\lambda|^2$ correspond two eigenmodes of opposite chirality, and the eigenvectors
of $D_{ov}(m=0)$ with complex conjugate eigenvalues
are superpositions of these two eigenvectors.
It is convenient to define the chiral projections ($P_\pm = \frac{1}{2}(1\pm \gamma_5)$)
so that the massive squared Hermitian overlap operator, with the usual
convention for the mass terms, is
\bee
H^2_\pm(m) =  P_\pm H^2(m) P_\pm=2(R_0^2-\frac{m^2}{4}) P_\pm (1\pm\epsilon(h))P_\pm+m^2 P_\pm \ .
\ee
Since the spectrum is doubled, the spectrum of one chiral sector of $H^2$ is equal to
the spectrum of a single quark flavor, apart from the real modes.
Their contribution can be included exactly~\cite{Bode:1999dd}.
Choosing to work in the chiral sector $\sigma$ which has no zero modes, the
partition function for a single flavor of quark is then
\bee
Z= \int [d\phi_\sigma^\dagger][d\phi_\sigma]\exp(-\phi_\sigma^\dagger H_\sigma(m)^2 \phi_\sigma
-|Q| \log(m/(2R_0)) ).
\ee
where $Q$ is the topological charge as defined by the number of zero modes of negative and
positive chirality, $Q=n_- - n_+$.
This system is then simulated by the Hybrid Monte Carlo algorithm~\cite{Duane:1987de}
with the modification of  Ref.~\cite{Fodor:2003bh} for topological boundaries.

As we have described in Ref.~\cite{ouralg}, the pseudofermion fields
are initialized with a set of chiral Gaussian random vectors $\xi_\sigma$ by
$\phi_\sigma=\sqrt{H_\sigma(m)^2} \xi_\sigma$. In our simulations we have used
two variations of this algorithm. For computing $\Sigma$, we have restricted the
simulation to fixed sectors of $Q$. To find the string tension and matching factor,
we have fixed the running chirality to be negative and restricted allowed topologies
to be $Q\le 0$. In the analysis of an ensemble generated with this algorithm, measurements
on the $Q=0$ configurations need to be reweighted with a factor $1/2$ compared to
those from configurations with non-trivial topology.

Our particular implementation of the Hybrid Monte Carlo algorithm has been
previously discussed in Refs.~\cite{DeGrand:2004nq,DeGrand:2005vb,Schaefer:2005qg}.
It uses multiple pseudofermions and stepsizes~\cite{Hasenbusch:2001ne,Urbach:2005ji}.

 We are using a planar kernel Dirac operator $d$ with nearest and
 next-to-nearest (``$\sqrt{2}$") interactions. 
To be precise,we parameterize the
associated massless free action by
\bee
S=\sum_{x,r} \bar \psi(x)\left[ \eta(r) + i \gamma_\mu \rho_\mu(r)\right] \psi(x+r)
\label{eq:fermionaction}
\ee
with  $r$ connecting nearest neighbors ($\vec r=\pm\hat\mu$;
$\eta=\eta_1$, $\rho_\mu = \rho_1$) and diagonal
neighbors ($\vec r=\pm\hat\mu \pm\hat\nu$, $\nu\ne\mu$;
$\eta=\eta_2$, $\rho_\mu= \rho_\nu = \rho_2$).
The constraint $\eta(r=0)=\eta_0 = -8\eta_1 -24 \eta_2$
enforces masslessness on the spectrum, and $-1 = 2 \rho_1 + 12\rho_2$ normalizes
the action to $-\bar \psi i \gamma_\mu \partial_\mu \psi$ in
the naive continuum limit. To speed up the code we require that each of the couplings
of a fermion to its neighbors is a projector, proportional to $1 \pm \hat n \cdot \gamma$.
This is a familiar trick for Wilson action simulations.  For nearest neighbors,
a projector action corresponds to the constraint $\eta_1 = \rho_1$ (up
to signs) and for the diagonal neighbors, $\eta_2 = \sqrt{2}\rho_2$.
The action we use in the simulations presented in this paper uses  $\rho_1 = -1/6$ and
 $\rho_2 = -1/18$.
We also add a clover term with the tree-level clover coefficient appropriate to
 this action of 1.278. 
We set the radius of the Ginsparg-Wilson circle $R_0$ to
1.2.

Our kernel operator $d$ is constructed from gauge links to
which three levels of isotropic stout blocking~\cite{Morningstar:2003gk} have
been applied. The  blocking parameter $\rho$ is set to 0.15.
 The sign function is computed using the
Zolotarev approximation
with an exact treatment of the low-lying eigenmodes $|\lambda\rangle$ of $h(-R_0)$.
We use the L\"uscher--Weisz gauge
action~\cite{Luscher:1984xn} with the  tadpole improved coefficients of
Ref.~\cite{Alford:1995hw}. Instead of determining the fourth root
of the plaquette expectation value $u_0=(\langle U_{pl}\rangle/3)^{1/4}$
self-consistently, we set it to 0.86 for all our runs as we did in our previous
publications.
We simulate on $10^4$ lattices at one value of the gauge coupling $\beta=7.7$
which we chose to be roughly at a lattice spacing of $0.16$ fm.
Our bare sea quark mass is  $am_q=0.05$.  We collected approximately 600
trajectories of data in fixed Q=0 and 1 sectors,
 and analyzed lattices from every fifth trajectory.

The use of three steps of stout blocking results in a considerable saving of computer time
compared to the two steps we have used previously.
This is seen by comparing the
number of inner Conjugate Gradient steps needed to evaluate the Zolotarev
approximation to the sign function.
In these simulations it is about 20.
With two steps, at essentially the same values of lattice spacing and quark mass,
it is about 47.

The range of smearing of the gauge fields for $N$ steps
stout blocking is~\cite{Bernard:1999kc} $\langle x^2 \rangle \sim 2 \rho a^2 N$.
 This corresponds to $\langle( x/a)^2\rangle^{1/2} \sim 0.77$ for $N=2$ and 0.95 for $N=3$
at $\rho=0.15$ such that also for the $N=3$ case the fermion action remains reasonably local.

\subsection{The lattice-regulated condensate}

We computed the lowest four eigenvalues $|\lambda|^2$ of the squared Dirac operator $H^2$.
They give us the eigenvalues of the overlap operator, which lie on a circle.
We apply the M{\"o}bius transform~\cite{Bietenholz:2003mi}
\bee
\tilde\lambda=\frac{\lambda}{1-\lambda/(2R_0)}
\ee
to project the eigenvalues onto the imaginary axis.

The data set  thus consists of a collection of eigenvalues, the
distribution of which is predicted by RMT as presented in 
Refs.~\cite{Damgaard:2000qt,Damgaard:2000ah}. It
 depends on one parameter $\Sigma$.

To avoid binning the data, which can introduce a bias, we
use the
Kolmogorov-Smirnov test~\cite{NR,Bietenholz:2003mi} as a measure for the goodness of the
fit. It compares the cumulative distribution function of the data
$C(x)$ to
the theoretical prediction $P(x)=\int_{-\infty}^x f(x) {\rm d} x$.
$C(x)$ is the fraction of eigenvalues  with a value smaller
than $x$.

The quantity of interest is the largest deviation of $P$ and $C$:
$D=\max_x |P(x)-C(x)|$. From this the confidence level is given by
\bee
Q_{KS}\left((\sqrt{N}+0.12+0.11/\sqrt{N})D\right)
\ee
with 
\bee
Q_{KS}(\lambda)= 2\sum_{j=1}^\infty (-)^{j-1}\exp(-2j^2\lambda^2).
\ee
In fits to a single eigenvalue distribution we maximize this quantity.
When fitting to more than one eigenvalue, we maximize the product over the individual
confidence levels.
 The errors on the fit parameter $\Sigma a^3$ are determined by the
bootstrap procedure.

\begin{figure}
\begin{center}
\includegraphics[width=0.58\textwidth,clip,angle=-90]{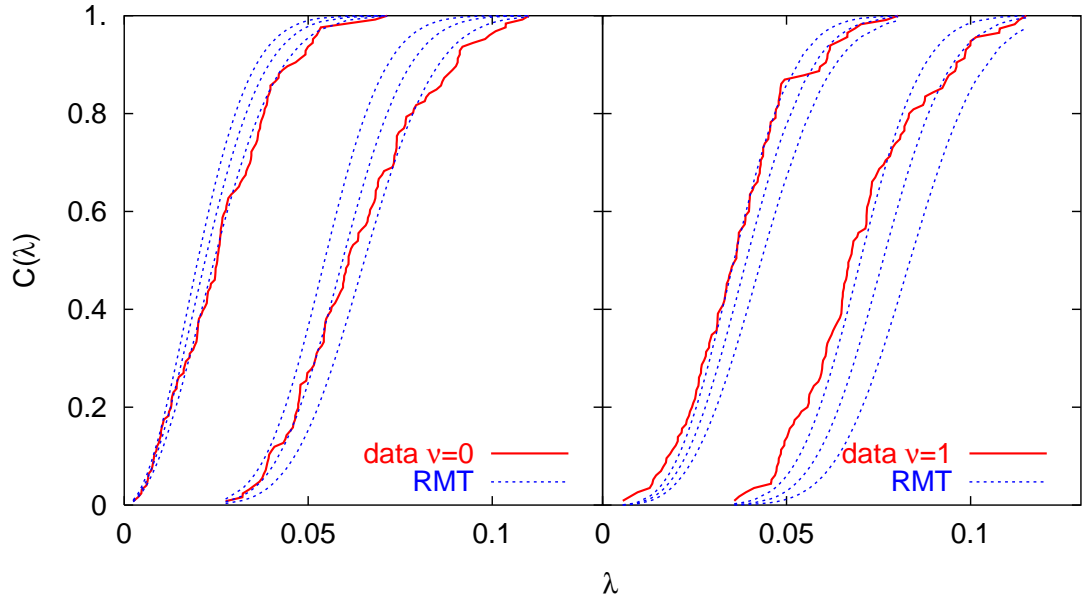}
\end{center}
\caption{
Cumulative distributions of the eigenvalues where
$C(\lambda)$ is the fraction of all gauge configurations where the n-th
eigenvalue is smaller than $|\lambda|$. We show the data for $n=1$ and $2$
for topological sector $\nu=0$ and 1. The dotted lines are the predictions
from RMT with the value of $\Sigma a^3$ obtained from  fits to the data 
(from left to right: fits C, B, A from Table~\ref{tab:KSsep}).
\label{fig:cm0.05}
}
\end{figure}

\begin{table}
\begin{tabular}{cc|cc|cc}
 Fit  & \ \  $\Sigma$ \ \ & $|\nu|$ \ \   &\ \ level 1\ \  &\ \ level 2\ \  \\
\hline
 \multirow{2}{*}{A}&   \multirow{2}{*}{0.0087(4)} & 0  & \fbox{0.87} & 0.02 \\
                   &                          & 1  & $2\cdot10^{-5}$& 0  \\
\hline
 \multirow{2}{*}{B}&   \multirow{2}{*}{0.0096(3)} & 0  & \fbox{0.04}    & 0.004   \\
                   &                          & 1  & \fbox{0.21}    & $1\cdot10^{-5}$ \\
\hline
 \multirow{2}{*}{C}&   \multirow{2}{*}{0.0102(4)}& 0  & $2\cdot10^{-4}$ & $2\cdot10^{-6}$ \\
                   &                          & 1  & \fbox{0.59}      & 0.21 \\
\hline
    
\end{tabular}
\caption{The confidence level of the individual distributions from the various fits.
The values for levels included in the fits are boxed.}

\label{tab:KSsep}
\end{table}

\begin{figure}
\begin{center}
\includegraphics[width=0.6\textwidth,clip,angle=-90]{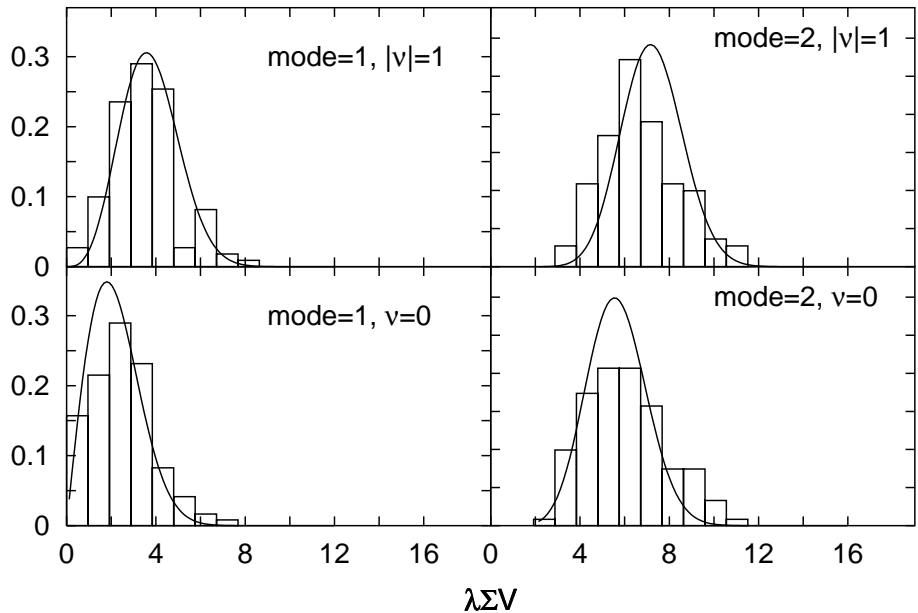}
\end{center}
\caption{Binned  eigenvalue distributions and the RMT prediction with $\Sigma a^3$ from
the fit to the lowest level in each topological sector (fit B from the table).
\label{fig:dist0.05}
}
\end{figure}

The results of our fits are presented in Table~\ref{tab:KSsep} and 
Figs.~\ref{fig:cm0.05}, \ref{fig:dist0.05}.
From a fit to the lowest level in $\nu=0$ alone (fit A) we extract $\Sigma a^3=0.0087(4)$ with a 
confidence level(CL) of 0.87. This means that the shape of the extracted curve is in very
good agreement with the theoretical prediction for the given statistics. The second level
in $\nu=0$ has a CL of 0.02, which is low but still acceptable.
A fit to the lowest level in $|\nu|=1$ (fit C) gives $a^3\Sigma a^3=0.0102(4)$ and a CL for
the lowest two level of 0.59 and 0.21. 
From a combined fit to the lowest level in each sector (fit B) we get $\Sigma a^3  =0.0096(3)$.
Even with the  extracted values of  $\Sigma a^3  $ seemingly apart, 
it is still possible that they are compatible with another, because
there is no correlation between the $\nu=0$ and $|\nu|=1$ ensembles.

Now that we have determined the optimal $\Sigma a^3$, the quality of the fits has to tell us
whether the RMT description of the data is valid. A possible source
for a deviation from the RMT prediction
is a too small volume.
Since we simulate at finite volume, we only expect the lowest few eigenvalues to match
the RMT curves. We observe that for the fits to one level alone, the fit to the second
eigenvalue 
in the same topological sector still
has an acceptable confidence level. The match between the $\Sigma$ in $\nu=0$ and  $|\nu|=1$
does not seem convincing. However, one has to keep in mind that the lowest 
two eigenvalues are calculated on the same gauge configurations and therefore are
correlated. On the other hand, there is no correlation between the extraction in the two
topological sectors. This can give rise to an apparent better match between the 
distributions of the two levels in one topological sector as compared to the match
between the lowest level in both sectors. 
Within our statistics, RMT seems to be applicable to the lowest eigenvalue
 in $\nu=0$ and $|\nu|=1$ and
the second lowest eigenvalue in $\nu=0$. Nevertheless, it is well possible that
the volume is still too small.

A main concern in any lattice calculation---and in particular ones using
new actions and algorithms---is the auto-correlation between consecutive
configurations. From our Markov chains, we saved every fifth configuration.
To find out whether this separation is enough, we used the following
technique: We split the data set into two sub-sets, one consisting of all
the even numbered
configurations and one of the odd numbered. A fit to the lowest eigenvalue in
$\nu=0$ and $|\nu|=1$ on the even and the odd numbered sample
gives $a^3\Sigma=0.0099(5)$ and 0.0094(4) respectively. The central values
are almost one sigma apart suggesting a weak correlation.

To make a more quantitative statement, we have computed $a^3 \Sigma$ on
matched Bootstrap samples (where if configuration $2i$ is part of the
''even'' Bootstrap sample, so is $2i+1$ in the odd sample). Correlations
between the two samples should show up as correlations between the
$\Sigma$ on the corresponding bootstrap samples. We therefore measured the
correlation matrix element (averaging over all bootstrap samples)
\bee
C=\langle (a^3 \Sigma_{\rm even}- a^3 \bar \Sigma_{\rm even})  (a^3 \Sigma_{\rm
odd}- a^3 \bar \Sigma_{\rm odd}) \rangle
\ee
and found $C/(a^6 \bar\Sigma_{\rm even}\bar\Sigma_{\rm odd})=0.0007(15)$ which is zero within errors.
Thus, we do not find any effects of auto-correlation beyond the five
trajectories by which we spaced our measurements.

In a second test of autocorrelations, we computed the integrated autocorrelation
time for individual eigenvalues from the data stream. We found integrated
autocorrelation times averaging about 1.5 (in units of the collection time,
5 trajectories), with an uncertainty of about 0.7. This again suggests a weak
correlation between successive measurements of eigenvalues.
From this analysis we conclude that the value of the condensate in lattice
units is $a^3\Sigma=0.0096(3)$.

\subsection{The length scale}
\begin{figure}
\begin{center}
\includegraphics[width=0.6\textwidth,clip]{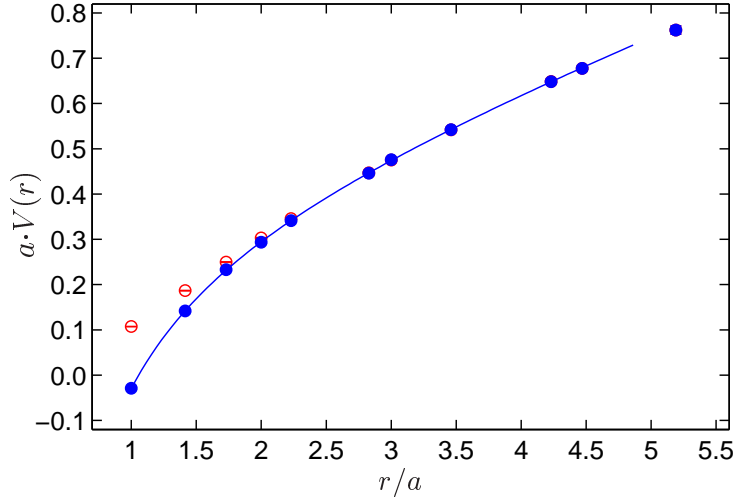}
\end{center}
\caption{The static quark potential in lattice units. The filled symbols
denote the potential after removing the artifacts introduced by the HYP
smearing.
\label{fig:potplot}
}
\end{figure}

As discussed above, an overall scale is obtained from the
static quark potential. The latter is extracted from the effective masses
of Wilson loops after one level of HYP smearing~\cite{Hasenfratz:2001hp,Hasenfratz:2001tw},
where the short-distance effects of
the HYP smearing are corrected using a fit to the perturbative lattice
artifacts. We measured the potential on 200 $8^3\times12$ configurations
at $am_q=0.05$ with the result shown in Fig.~\ref{fig:potplot}. From the
parameters of the fit we obtain
$r_0/a=3.37(10)$ for the Sommer parameter (\ref{above_equation}) and
$a\sqrt{\sigma}=0.318(25)$ for the string tension.
With $r_0=0.5$ fm, this gives a lattice spacing of $a=0.15$ fm.

\subsection{Matching lattice and continuum regularizations}

A matching factor is needed to convert the lattice
calculation of the quark condensate to its corresponding $\overline{\rm MS}$
value.
To get the matching factor, we use the RI' scheme introduced in
Ref.~\cite{Martinelli:1994ty}, and we follow the procedure described in
Ref.~\cite{DeGrand:2005af}, in which we calculated the matching factors for a
quenched simulation.  The RI' scheme results
in the chiral limit can be
converted to $\overline{\rm MS}$ values at $\mu=2$ GeV by using the ratios
connecting the two schemes. The ratios were computed by continuum
perturbation theory to three loops~\cite{Franco:1998bm, Chetyrkin:1999pq}.

These simulations should not be restricted in topological sectors.
To produce a matching factor, one needs simulations with a momentum scale
short enough to be free from nonperturbative effects and yet not so short as
to be affected by discretization errors.
Our data set consisted of 57 $8^4$ configurations from about 300 trajectories
at $am_q=0.05$.

 The $8^4$ lattice is periodic in space directions and
antiperiodic in the time direction. Therefore the momentum values are
\begin{equation}
ap_\mu=\left(\frac{2\pi}{8}k_x,\frac{2\pi}{8}k_y,\frac{2\pi}{8}k_z,
\frac{\pi}{8}(2k_t+1)\right).
\end{equation}
We choose the values of $k_\mu$ such that
the momentum values lie
as close as possible to the diagonal of the Brillouin zone. The maximum
value of $ap=2.256$ corresponds to $k_\mu=(2,1,1,1)$. The quark
propagators are cast from a point source and then projected
to the desired momentum values.

\begin{figure}
\begin{center}
\includegraphics[width=0.6\textwidth,clip]{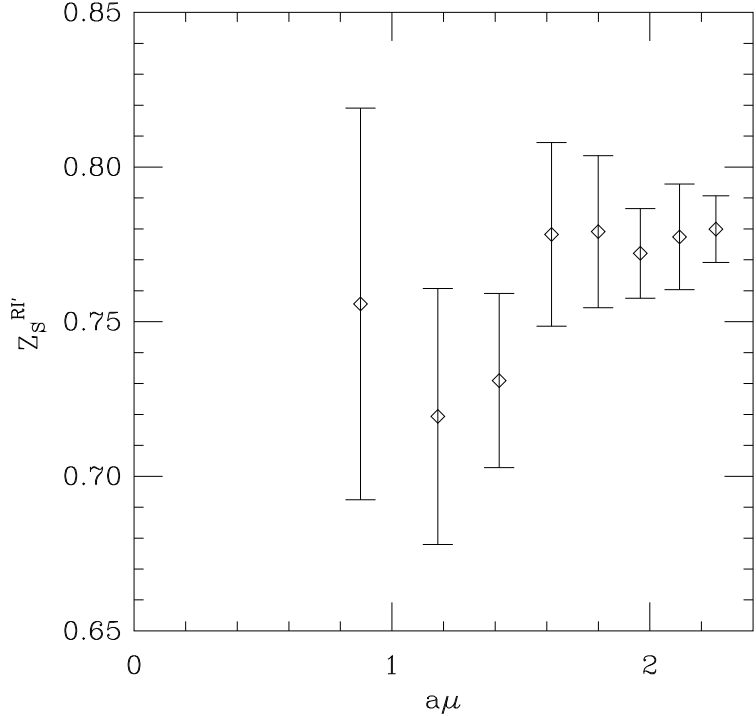}
\end{center}
\caption{$Z_S^{\rm RI'}$ for the one flavor overlap simulation.}
\label{fig:zs}
\end{figure}

$Z_S^{\rm RI'}$ for the scalar density is shown in
Fig.~\ref{fig:zs}.
The values of $Z_S^{\rm RI'}$ are listed in
Table~\ref{zstable}.  From our lattice spacing determined from the Sommer
parameter, $\mu=2$~GeV corresponds 
to $a\mu=1.504$.
 The 2 GeV RI' value is obtained from a
linear interpolation from the two closest $a\mu$ points of the data. The
result is $Z_S^{\rm RI'}(2~{\rm GeV})=0.75(3)$.
\begin{table}
\caption{Values of $Z_S$ in the RI' scheme.
The inverse lattice spacing is 1.330 GeV
from the Sommer parameter. Therefore $\mu=2$~GeV corresponds to
$a\mu=1.504$. The value at this point is obtained from a
linear interpolation from the two closest $a\mu$ points of the data.}
\begin{center}
\begin{tabular}{cllll}
\hline\hline
$a\mu$     &0.878  &1.178  &1.416  &1.619\\
$Z_S^{RI'}$&0.76(6)&0.72(4)&0.73(3)&0.78(3)\\
\hline
$a\mu$     &1.800  &1.963    &2.115    &2.256\\
$Z_S^{RI'}$&0.78(2)&0.772(15)&0.777(17)&0.780(11)\\
\hline\hline
\end{tabular}
\label{zstable}
\end{center}
\end{table}

The conversion ratio from the RI' to the $\overline{\rm MS}$ scheme for the scalar
and pseudoscalar densities, from~\cite{Franco:1998bm,Chetyrkin:1999pq}, needs an $\alpha_s$.
In Landau gauge and to three loops, and for one flavor,
the ratio is given by
\begin{eqnarray}
\frac{Z_S^{\overline{\rm MS}}}{Z_S^{\rm RI'}}&=&
\frac{Z_P^{\overline{\rm MS}}}{Z_P^{\rm RI'}}
=1+\frac{16}{3}\frac{\alpha_s}{4\pi}+\left(\frac{1375}{6}
-\frac{152\zeta_3}{3}\right)\left(\frac{\alpha_s}{4\pi}\right)^2
\nonumber \\
&&+\left(\frac{32149271}{2916}-\frac{215489\zeta_3}{54}
-\frac{80\zeta_4}{3}
+\frac{2960\zeta_5}{9}\right)\left(\frac{\alpha_s}{4\pi}\right)^3
+O(\alpha_s^4),
\label{zspratiof1}
\end{eqnarray}
where $\zeta_n$ is the Riemann zeta function evaluated at $n$.

To get numerical results of the above ratio, we
use the coupling constant from the so-called ``$\alpha_V$" scheme.
As in the appendix of Ref.~\cite{DeGrand:2005vb}, from the one-loop
expression relating the plaquette to the coupling
\begin{equation}
\ln\frac{1}{3}\Tr U_p=-\frac{8\pi}{3}\alpha_V(q^*)W,
\end{equation}
where $W=0.366$ and $q^*a=3.32$ for the tree-level L\"uscher Weisz
action, we obtain $\alpha_V(3.32/a)=0.173$. Then $a\Lambda_{\overline{MS}}$ is
calculated and $\alpha_s^{\overline{\rm MS}}$(2 GeV) is determined by using
$\beta_0=31/12\pi$ and $\beta_1=268/48\pi^2$ for one flavor QCD.
We find $\alpha_s^{\overline{\rm MS}}$(2 GeV)$=0.194$. Substituting
$\alpha_s^{\overline{\rm MS}}$ into Eq.~(\ref{zspratiof1}), we get
$Z_S^{\overline{\rm MS}}/Z_S^{\rm RI'}=1.147$ and therefore $Z_S^{\overline{\rm
MS}}(2~{\rm GeV})=0.86(3)$.

\section{Results \label{sec:results}}
From Sec. \ref{sec:method} the continuum-regularized condensate is
\beea
r_0^3 \Sigma(\overline{MS},\mu=2 \ {\rm GeV}) &=&
 Z_s(\mu,a) \times  \Sigma a^3 \times (\frac{r_0}{a})^3 \nonumber \\
 &=&
 0.86(3)
 \times
 0.0096(3)
 \times
 (3.37(10))^3 \nonumber \\
&=& 0.317(32).
\eea
Taking the real-world value for $r_0=0.5$ fm, this is
\bee
\Sigma(\overline{MS},\mu=2 \ {\rm GeV}) = 0.0194(20){\rm GeV}^3
\ee
or
\bee
( \Sigma(\overline{MS},\mu=2 \ {\rm GeV}) )^{1/3} = 0.269(9) {\rm GeV}.
\ee

\section{Conclusions \label{sec:conclude}}
Our result is in remarkable agreement with the central value of Eq.~(\ref{eq:asv}).
Thus at least for the condensate the $1\pm1/N_c$ estimates of the correction
to the large $N_c$ result, the extreme values of Eq.~(\ref{eq:asv}), seem to be quite
pessimistic.
In view of this finding more predictions from large $N_c$ orientifold QCD might
be interesting even without the knowledge of the subleading corrections.

A summary of previous calculations of the condensate has recently been given
 by McNeile~\cite{McNeile:2005pd}.
Comparing the quenched
determinations there, a three-flavor
prediction by McNeile, and our result,
the condensate seems to be a quantity which is not very $N_f$ dependent.

Other tests of orientifold equivalence will be difficult. To check
the prediction of Ref.~\cite{Armoni:2003fb} that
$m_{\eta'}^2/m_\sigma^2=1+O(1/N_c)$ is nontrivial because it requires
disconnected diagrams. A similar degeneracy of hybrids described in
Ref.~\cite{Gorsky:2004nb} will be hard because the sources for ordinary hybrids
 are noisy in QCD.

\section*{Acknowledgments}

This work was supported in part by the US Department of Energy.
T.D. would like to thank M. Strassler for conversations.

\end{document}